\documentclass[sigconf,review,anonymous]{llncs}

\usepackage{graphicx}
\usepackage{amsfonts}
\usepackage{amssymb}

\usepackage{color}
\usepackage{caption}
\usepackage{subcaption}
\usepackage{listings}
\usepackage{mathtools}
\usepackage{float}
\usepackage{hyperref}
\usepackage{wrapfig}
\usepackage{todonotes}

\usepackage{orcidlink}
\usepackage{algorithm}
\usepackage[noEnd=true,rightComments=true,commentColor=black,beginComment=//~,italicComments=false]{algpseudocodex}

\usetikzlibrary{arrows, decorations.pathmorphing, decorations.pathreplacing}

\usepackage{macro}

\newcommand\simgrid[1]{\textsc{#1}}

\newlength\myindent
\begin{document}
\title{Towards Efficient Verification of Parallel Applications\\ with Mc~SimGrid}

\author{Mathieu Laurent\orcidlink{0009-0003-5997-2699},
  Thierry J\'eron\orcidlink{0000-0002-9922-6186}
  and Martin Quinson\orcidlink{0000-0001-7408-054X}}

\institute{Univ Rennes, Inria, CNRS, Irisa, Rennes, France}

\maketitle
\begin{abstract}
    Assessing the correctness of distributed and parallel applications is notoriously difficult due to the complexity of the concurrent behaviors
    and the difficulty to reproduce bugs.
    In this context, Dynamic Partial Order Reduction (DPOR) techniques have proved successful in exploiting concurrency to verify applications without exploring all their behaviors.
    However, they may lack of efficiency when tracking non-systematic bugs of real size applications.
    In this paper, we suggest two adaptations of the Optimal Dynamic Partial Order Reduction (ODPOR) algorithm with a particular focus on bug finding and explanation.
    The first adaptation is an out-of-order version called RFS ODPOR which avoids being stuck in uninteresting large parts of the state space.
    Once a bug is found, the second adaptation takes advantage of ODPOR principles to efficiently find the origins of the bug.
\end{abstract}

\section{Introduction}
% \color{blue}
% \begin{itemize}
% \item On cherche des bugs dans des codes MPI et/ou multi-threadÃ©s
% \item deadlock, fautes syst\`emes (segfaults ou autre), et assertion failures
% \item D\'efinir ce qu'on fait : dynamic software model checking. Discussion sur stateful/stateless (on garde des metadonn\'ees mais pas l'\'etat)
%   test exhaustif d'un programme pour un input donn\'e, en utilisant des techniques de r\'eduction venant du MC
% \item Ref a Dirk Bayer. SMC: 20 years and beyond. Pour utiliser le meme vocabulaire qu'eux
%   un paragraphe de sota en intro
% \end{itemize}
% \color{black}

We consider parallel applications made of interacting actors that can share the
memory of a single computer, or can be geographically distributed across the network.
These actors interact through message passing over the network, or through classical synchronization mechanisms such as mutexes or semaphores.
This class of applications encompasses a large amount of real-world programs, such as
message-passing ones that are commonly built upon the MPI standard~\cite{MPI} in
High-Performance Computing, or classical multithreaded applications built for example upon the POSIX \texttt{pthread} standard~\cite{pthreads96}.
These applications are notoriously difficult to get right as they add the challenges of
concurrent and multithreaded applications (race conditions, deadlocks, livelocks) to the
challenges linked to asynchronous message passing (communication mismatch), and memory
coherency in distributed settings.

%% Parallel distributed applications are complex software, whose correction is a challenge.
%% The main difficulties come from the absence of a global clock, distribution of the memory and concurrency between actions of different processes (e.g. race conditions).

The main challenge to the correctness of these applications lays in the fact that the
ordering of events changes the behavior. Moreover, triggering existing bugs also becomes challenging in such complex applications.
The purpose of our work is therefore to design automatic verification methods that are both
complete for a fixed input, and efficient for finding bugs in applications whose behaviors
are supposed to be finite but possibly non-deterministic.

Testing can be efficient for detecting bugs in software. It can be made formal, in particular by synthesizing tests and their verdicts
automatically from a  model of the system. Unfortunately, it suffers from
non-completeness and its difficulty to reproduce bugs for distributed applications.
Among formal verification techniques, model checking is complete.
In its classical form it requires building a formal model by abstraction of the system behavior.
Completeness is then relative to this model, not necessarily the real system.
Moreover, it often does not scale to real size programs, which hampers completeness
in practice. Software model checking is an alternative which focuses on the software code
itself, releasing the user from the burden of building a formal model. This approach
encompasses a set of techniques inspired from static analysis, model checking or automated
proving, and proved to be scalable~\cite{SMC_Rupak}.

Among these, dynamic partial order reduction
(DPOR)~\cite{DPORGodefroid} comprises techniques that are complete (under some
hypothesis), and exploit the independence
of actions of concurrent actors to avoid explicitly traversing all behaviors.
DPOR is stateless, meaning that system states are not stored, since they are too large.
In the last decade, new developments proposed optimal
solutions (see \eg~\cite{ODPOR-POPL14,SDPOR-JACM17,UDPOR-Concur15}). In the same time,
those techniques have been tuned for relaxed memory models where even local actions may be
independent (see \eg~\cite{zhang2015dynamic,Koko-POPL22}).

Our goal is to verify real parallel programs, for which we do not dispose of a model, but can execute their code. 
We build upon Mc~SimGrid~\cite{McSimGrid}, a stateless model checker
targeting parallel distributed applications that interact through message passing or
through synchronization mechanisms (mutex, barrier, etc.).
We use this tool as an
experimental platform for the study of model checking techniques in this domain.Underneath, Mc~SimGrid leverages the SimGrid
simulator~\cite{SimGrid} to observe and control the execution of the applications under
scrutiny.

In this paper, we propose adaptations of the ODPOR algorithm~\cite{ODPOR-POPL14}.
We want to improve the practical efficiency of ODPOR when searching for bugs. Indeed, even if ODPOR
is complete and optimal, its depth-first search (DFS) nature may make it inefficient at
detecting bugs by forcing the full exploration of uninteresting parts of the state space
before reaching a bug.
Our first contribution propose a variant of ODPOR based on an out-of-order traversal that 
increases the chance to detect bugs quickly.
We also aim at increasing the usefulness of counter-examples when bugs are found,
when the usually long sequences of actions reveal hard to analyze by humans.
We define a notion of \newdef{critical transition}, which separates correct from incorrect executions in an execution path to a bug.
It is thus part of the bug root cause.
Our second contribution  is to further adapt the ODPOR algorithm so that, when a bug is found, it efficiently searches this critical transition.
All algorithms are implemented in the prototype Mc~SimGrid and experimented on some
benchmarks. The analysis of the experiments show promising results.

The rest of this article is organized as follows. Section~\ref{sec:background} provides
some background information about the considered class of applications, the limits of
the programming model, and how classical DPOR techniques are leveraged to verify real
distributed applications. Section~\ref{sec:contrib} presents the contributions of this
work, while Section~\ref{sec:xp} provides some preliminary experiments to evaluate these
contributions. Section~\ref{sec:cc} concludes this paper by sketching possible future
works.

\section{Adapting Model Checking Techniques to the Verification
  of Parallel Applications}\label{sec:background}

In this section, we first sketch the programming model for parallel applications implemented in Mc~SimGrid.
We then introduce the formalism necessary for DPOR algorithms,
and then explain the principles of the ODPOR algorithm.

\subsection{Extended Programming Model}\label{subsec:progmodel}
\label{subsec:prog-model}
  
%% \color{blue}
%% \begin{itemize}
%% \item MPI comme dans le papier precedent
%% \item Threads (sthread fonctionne -- sauf race conditions), mutex sem condvar barriere,
%% \item Wait/Test sur Any/All => ca donne du non-determinisme, version gentille et gerable
%% \item Iprobe?
%% \item RMA (sauf race conditions) ?
%% \end{itemize}
%% \color{black}

Real parallel systems can be built with many programming interfaces, each containing many
primitives. The amount of primitives makes it impractical to reason at that level to
verify the resulting applications~\cite{GaneshAllMPIPrimitives}. Instead, we build upon an
abstract model entailing only about 20 actions. This reduced set of primitives are used at
the core of the SimGrid simulator to mediate all interactions between interacting
actors\footnote{We use actor and process interchangeably to designate an entity
  interacting with others either by message passing or shared memory},
with the exception of direct memory sharing between actors. Over the years,
this set of primitives was proven sufficient to represent a large class of parallel
applications: \cite{SMPI} presents a SimGrid implementation of the MPI
standard~\cite{MPI}, that prevails in High-Performance Computing. \cite{sthread} presents
a SimGrid implementation of the POSIX \texttt{pthread} library, that is used by
multithreaded UNIX applications. These near-complete reimplementations of the standards
allow most of the applications written with these interfaces to run unmodified on top of
SimGrid.
%thus proving that the Mc~SimGrid programming model encompasses a large class of real applications.
% making it theoretically possible to verify these applications with our system.

This programming model is composed of several subsystems, each proposing specific
actions allowing actors to interact through shared objects:
The \emph{network subsystem} proposes \textit{mailbox objects} allowing receiving actors
to be matched with sending ones. The corresponding actions are asynchronous and
non-blocking to meet the needs of the MPI interface: \simgrid{async\_send} and
\simgrid{async\_recv} (receive) represent the start of a background communication. The
usual asynchronous communication pattern becomes the combination of an
\simgrid{async\_send} and of an \simgrid{async\_recv} immediately followed by a
\simgrid{wait}. This extension is mandatory to precisely model the MPI semantics as with
the \texttt{MPI\_Isend}, \texttt{MPI\_Irecv} and \texttt{MPI\_Wait} primitives.
The \emph{synchronization subsystem} proposes four classical types of objects:
\textit{mutex, semaphore, barrier,} and POSIX \textit{condition variable}. The
corresponding actions implement classical operations such as \texttt{lock/unlock},
\texttt{test} and \texttt{signal/broadcast} for condition variables. Although the POSIX
standard describes only synchronous versions of these operations, we refine them into an
asynchronous action followed by a \simgrid{wait} for symmetry with the network subsystem.
Finally, \simgrid{waitall} can be used to wait atomically for the completion
of all actions in a given set while \simgrid{waitany} blocks until the completion of an
arbitrary action among a given set.
This last action is \textit{non-deterministic} in the sense that an actor executing it returns
\textit{any} of the terminated actions in the set. Similarly, \simgrid{random} can be used
to explore several alternate outcomes. For example, \simgrid{random(\{1, 2\})} can return
either 1 or 2.

% On parle pas de collectives MPI: les détails sont dans le papier SMPI

% Actions cachées sous le tapis:
% * COMM_IPROBE (trop spécifique à MPI et embrouillé à expliquer)
% * ACTOR_JOIN/ACTOR_SLEEP (spécifique à pthread et pas super passionnant)
% * le fait que les WAIT sont spécifiques à l'action attendue pour rendre certaines
%   indépendances statiques (genre SEM_WAIT et MUTEX_UNLOCK)
% * COMM_TEST n'est pas détaillé et on insiste pas sur le fait qu'on peut tester
%   certains objets mais pas tous (BARRIER_TEST n'existe pas, par exemple)

%RANDOM, ACTOR_JOIN, ACTOR_SLEEP,
%TESTANY, WAITANY,
%BARRIER_ASYNC_LOCK, BARRIER_WAIT,
%COMM_ASYNC_RECV, COMM_ASYNC_SEND, COMM_IPROBE, COMM_TEST, COMM_WAIT,
%MUTEX_ASYNC_LOCK, MUTEX_TEST, MUTEX_TRYLOCK, MUTEX_UNLOCK, MUTEX_WAIT,
%SEM_ASYNC_LOCK, SEM_UNLOCK, SEM_WAIT,
%CONDVAR_ASYNC_LOCK, CONDVAR_BROADCAST, CONDVAR_SIGNAL, CONDVAR_WAIT,

%\smallskip
This programming model  refines the one described in \cite{Pham19a}, enriched with the \texttt{pthread} semantics.
It is probably sufficient to encode the semantics
of other libraries such as the BSD sockets that base almost all Internet communications.
The main limitation of this model is that it does not capture direct memory
accesses. Memory reads and writes are not observable in our model for sake of
verification efficiency. This implies that data races, for instance, are out of the scope
of our study. Instead, we focus on bugs triggered by specific orderings of
actions, such as deadlocks or assertion errors due to communication mismatch or
synchronization misuses. Previous work have shown the importance of this class of bugs in
real applications~\cite{locksMisuses}.

\subsection{Notations and Definitions}\label{subsec:notations}

Formally, we consider a parallel system composed of $n$ programs
$p_1, \ldots, p_n \in\mathbb{P}$. The global behaviour of this system can be described by
a \newdef{labelled transition system} (LTS) $\mathcal{M} = (S, \Act, s_0, \mathcal{T})$ where $S$
is a  \newdef{set of states} encoding local states of processes, contents of communication
channels and status of synchronization objects, $s_0\in S$ is the \newdef{initial state} where each process
$p$ is in its initial state, communication channels are empty and synchronization objects
free, $\Act=\bigcup_{i=1}^n \Act_i$ is the alphabet of \newdef{actions}, the disjoint union of local
alphabets, $\mathcal{T}\subseteq S \times \Act \times S$ is the 
\newdef{transition relation}, each \newdef{transition} 
$(s,a,s')\in\mathcal{T}$ sometimes written $\transition{s}{a}{s'}$. 
We note $\proc((s,a,s')) = \proc(a) = p_i$ if $a \in \Act_i$
meaning that $\transition{s}{a}{s'}$ is executed by process $p_i$. We denote by
$\enabled(s)$ the set of transitions that can be taken from state $s$. In the following,
we assume that actors are \textit{deterministic}, \ie, for
each $s\in S$, if $t_1, t_2\in\enabled(s)$, then $\proc(t_1)\neq\proc(t_2)$. This allows
one to identify the transitions in a given state by their actors. Our results and
implementation hold for non-deterministic actors, but distinguishing the transitions
outcomes of a given actor in a given state would make the notations artificially complex.

We restrict our study to acyclic LTS.
An \newdef{execution}  of $\mathcal{M}$ is a finite sequence
of transitions starting in the initial state 
\(E := t_1\cdot t_2\cdots t_n = (s_0, a_1, s_1)\cdot(s_1, a_2,
s_2)\cdots(s_{n-1},a_n,s_n) \in\mathcal{T}^*\) also 
written $\transitions{s_0}{a_1\cdot a_2\cdots a_n}{s_n}$.
With the remark above, $E$ can also be written by its
sequence of processes $\proc(t_1)\cdots\proc(t_n)$.
$E$ is \newdef{maximal} when $\enabled(\last(E)) = \emptyset$.
The following notations are introduced:
\begin{itemize}
\item $\dom(E) := \{1,\ldots,n\}$, the range of transitions in $E$,
\item $E_i := t_i$ for $i \in \dom(E)$, the $i^{th}$ transition of $E$,
\item $E\cdot t := t_1\cdots t_{n}\cdot t$, for $t = (s_n, a_{n+1}, s_{n+1})$, the
  concatenation of $E$ with $t$,
\item $E\vdash t$ (or equivalently $E\vdash p$ where $p=\proc(t)$), for $t \in T$, the fact that $E\cdot t$ is a valid execution,
\item $\sub(E, t_i) := t_1\cdots t_{i}$ for $i\in \dom(E)$, the prefix of $E$, up to $t_i$
  included,
\item $\pre(E, i) := s_{i-1}$ for $i\in \dom(E)$, the state reached before executing $t_i$,
\item $\last(E) := s_n$, the last state of the execution $E$.
\end{itemize}

For sake of algorithmic
simplicity, we  assume a property called \textit{persistency}:
an enabled process cannot be disabled before it executes some action.
Formaly, for any execution $E$, any process $p\in\mathbb{P}$,
if $E\vdash p$, then for any other process $q \neq p$,  $E\vdash q$ implies $E\cdot q \vdash p$.
Notice that in the programming model, persitency justifies to split blocking calls such as \simgrid{mutex\_lock} into a
non-blocking \simgrid{mutex\_async\_lock}, that adds the caller's ID to the list of actors requesting this mutex,
followed by a blocking \simgrid{mutex\_wait}, that becomes enabled only
when the caller's ID is first in that list. While an atomic \simgrid{mutex\_lock} would
not be persistent (it gets disabled if another actor gets the lock), both
\simgrid{mutex\_async\_lock} and \simgrid{mutex\_wait} are actually persistent.

\subsection{State Space Reduction using ODPOR}
\label{subsec:DPOR}

%% \color{blue}
%% \begin{itemize}
%% \item Principe de base: relation d'independance/dependance, calculee sur notre modele de
%%   calcul avec qques exemples
%% \item En toute generalite il faut l'independance entre toutes les operations impliquees
%%   (papier Ganesh TLA collectives MPI), mais nous on simplifie le pb en adaptant le modele
%%   de calcul.
%% \item Pour les comms, on utilise que les asynchrones et les collectives sont executees
%%   selon les vrais algos.
%% \item De plus, on utilise la persistance du modele de calcul pour optimiser l'algo de
%%   reduction
%% \item On donne l'intuition de pourquoi cette reduction est optimale
%% \item on ne parle pas trop des autres algos (SDPOR etc) car on veut pas les comparer dans
%%   cet article
%% \end{itemize}
%% \color{black}

Dynamic partial order reduction (DPOR)~\cite{DPORGodefroid} is a software model
checking technique that exploits the independence between concurrent actions.
To avoid the exploration of the full state space, it relies on the equivalence of executions by commutation of adjacent transitions carrying independent
actions.

Intuitively, two actions are independent if firing one cannot enable or disable the other,
and they commute, \ie their execution order does not impact the final
result.
Formaly, a valid \newdef{independency} relation for $\mathcal{M}$
is an irreflexive and symmetric relation $\mathcal{I}\subseteq \Act\times\Act$
such that, for any state $s\in S$ and any pair of \newdef{independent} actions $(a_1, a_2)\in\mathcal{I}$, we have:
\begin{itemize}
\item if there exists $s'\in S$ such that $\transition{s}{a_1}{s'}$, then
  $a_2 \in\enabled(s)$ if and only if $a_2\in\enabled(s')$,
\item if $a_1, a_2\in\enabled(s)$, then there exists $s'\in S$ such that
  $\transitions{s}{a_1a_2}{s'}$ and $\transitions{s}{a_2a_1}{s'}$.
\end{itemize}

The \newdef{dependency} relation $\mathcal{D} = \Act\times\Act \setminus \mathcal{I}$ is the complementary
relation. In this paper we only consider a statically defined valid dependency relation,
\ie, based on the semantics of the programming model of Mc~SimGrid. For instance, we can state
that \simgrid{async\_send} and \simgrid{async\_recv} actions are always independent, while two
\simgrid{mutex\_lock}   concerned by the same
mutex are dependent~\cite{Pham19a}.

\begin{figure}[thb]
  \begin{center}
\begin{lstlisting}[language=C, frame=single,  captionpos=b,  xleftmargin=0.001\textwidth, xrightmargin=0.001\textwidth,
  basicstyle=\footnotesize\ttfamily]  % Start your code-block

P1               P2              P3           
ASYNC_SEND(P3)   ASYNC_SEND(P3)  ASYNC_RECV(from any); WAIT()
                                 ASYNC_RECV(from P2) ; WAIT()
\end{lstlisting}                      
  \caption{A communication pattern using the \emph{any} wildcard.}
  \label{lst:pattern}
  \end{center}
\end{figure}

For a sequence of transitions  $E = t_1\cdots t_n$, the \newdef{happens-before}~\cite{LamportHB} relation
$\to_E$ is the smallest transitively closed relation on $\dom(E)$ such that
if $i<j$ and the actions carried by $t_i$ and $t_j$ are dependent then $i\to_E j$.
\begin{example} Consider the execution $E= P_1\cdot P_2\cdot P_3\cdot P_3$ from
  Fig.~\ref{lst:pattern} (remember that a transition can be identified with its actor).
  We have that $E_1 \to_E E_2$ because two \simgrid{async\_send} actions
  to the same actor are dependent, and $E_3\to_E E_4$ because actions from the same actors
  are causally ordered.
\end{example}

Let $\simeq$ be the equivalence relation on executions such that $E \simeq E'$ if $\dom(E) = \dom(E')$
and $\to_E = \to_E'$.  
A \newdef{Mazurkiewicz's trace}  $[E]_\simeq$, 
is the class of executions equivalent for $\simeq$ and containing $E$.

DPOR algorithms aim at exploring the smallest number of executions while ensuring soundness of the
exploration. This is ensured by exploring at least one execution per Mazurkiewicz's
trace. To that extend, reduction algorithms aim at finding \newdef{reversible races}, \ie,
identifying pairs of dependent transitions which inversion leads to a distinct
Mazurkiewicz's trace. Formally,  $t_i$ and $t_j$ with $i<j\in\dom(E)$ are in
reversible race, noted $i\precsim_Ej$, when:
\begin{itemize}
\item $\proc(t_i)\neq \proc(t_j)$, and  $i \to_E j$,
\item there is no $k$ such that $i<k<j$, $i \to_E k$ and $k \to_E j$,
\item for any $E' \in [E]_\simeq$ of the form $E'=F\cdot t_i\cdot t_j\cdot F'$, then
  $F\vdash t_j$.
\end{itemize}
The first condition states that $t_i$ and $t_j$ belong to different actors and are
dependent, ensuring that their inversion produces a different Mazurkiewicz's trace. The
second condition ensures that there is no intermediate transition that would be in
reversible race with both of them. The last one ensures that $t_i$ does not enable $t_j$.
We then get that if $i\precsim_Ej$, there exists $E'\not\simeq E$ with $j\to_{E'}i$.
The satisfaction of properties such as absence of deadlock or
assertion violation is consistent with Mazurkiewicz's traces~\cite{DPORGodefroid}, 
meaning that for any such property $\varphi$, if $E \simeq E'$ then
$\varphi\models E \iff \varphi\models E'$. The principle of a DPOR algorithm is then to
verify those properties by building a so called ``reduced LTS'', where each Mazurkiewicz's trace is
represented by at least one execution.

Optimal Dynamic Partial Order Reduction (ODPOR)~\cite{ODPOR-POPL14} is a  recent
DPOR technique.
Like other DPOR algorithms, ODPOR is based on a stateless depth-first exploration of a reduced state space.
The exploration also uses a set of processes called \newdef{sleep set} 
to avoid exploring executions equivalent to already traversed ones. 
The original DPOR algorithm stores in each traversed execution $E$ a \newdef{persistent set},
\ie~ a set of processes that should be visited later when backtracking,
populated when a reversible race just after $E$ is detected in some continuation execution.
The optimality of ODPOR is gained by associating to each traversed execution $E$ a more precise information called \newdef{wakeup tree},
a set of initial sequences that should be visited later when backtracking to $E$.

The crucial notion of wakeup tree is detailed now.
First, we define \newdef{weak initials} of a sequence $w$ after a prefix
$E$: $\weakinitial{E}(w)$ is the set of processes $p$ \st~there exists sequences $w'$ and $v$ satisfying
$E\cdot w\cdot v \simeq E \cdot p \cdot w'$. Intuitively, $\weakinitial{E}(w)$ are the
processes in $\enabled(E)$ that have no happens-before predecessor in $w$.
Let an \newdef{ordered tree} $(B, \prec)$ be a prefix-closed set of executions, each {\em node} being an execution and the {\em root} is the empty one $\emptyseq$.
A \newdef{wakeup tree} after a prefix $E$ written $\wut(E)$, relative to a set of processes $P$ (the sleep set in $E$) is then an ordered tree $(B,\prec)$ \st:
\begin{itemize}
\item for every leaf node $w\in B$, $\weakinitial{E}(w) \cap P =\emptyset$,
\item for every node $u.p$ and leaf $u.w$ in $B$, if $u.p\prec u.w$ then $p\notin\weakinitial{E.u}(w)$.
\end{itemize}

Both conditions are necessary to achieve  optimality of ODPOR,  meaning that each Mazurkiewicz's trace is explored exactly once by the algorithm
and no sleep-set-blocked (SSB) execution is visited
(an SSB execution is an execution equivalent to an already explored one, but discovered lately when already partially explored).
In fact, if $P$ above is the sleep set computed when backtracking to $E$, the first condition ensures that no SSB will be visited.
The second condition prevents inserting a sequence for which a possibly equivalent prefix already exists in a given wakeup tree.
We skip most details here, and report to~\cite{ODPOR-POPL14} for the explanations on how to
construct wakeup trees and the corresponding insertion and suppression operations.
Simply remember that  $\wut(E)$ is populated each time a reversible race at $E$ is detected on a complete continuation of $E$,
and consists in inserting in order an initial execution sequence needed to reverse the race from $E$. 

\begin{example}
  The example shown in Fig.~\ref{lst:pattern} contains a deadlock: if the \simgrid{async\_send}
  from $P_2$ is executed first and matches the \simgrid{async\_recv}(from any) of $P_3$ then the
  second \simgrid{async\_recv} of $P_3$ can never be fulfiled, and process $P_3$ can never make
  progress. On this simple example, both DPOR
  and ODPOR behave the same. They explore a first execution, \eg $P_1\cdot P_2\cdot P_3\cdot P_3$. After
  discovering a reversible race between the first and the second transitions (the two
  \simgrid{async\_send}), they try an execution with those two transitions reversed. While the
  DPOR will only force the execution of $P_2$ before $P_1$ (at the start of the program),
  ODPOR will force the whole sequence $P_2\cdot P_3\cdot P_3$ also before $P_1$.
  By further restricting  the degree of freedom in the exploration, ODPOR achieves optimality. 
\end{example}

 Applying ODPOR to our programming model requires defining a valid static dependence relation~\cite{Pham19a}.
 It consists in a set of boolean functions,
 one for each pair of action types, that are used extensively during the ODPOR exploration of the application.
 The number of functions being quadratic in the number of actions types,
 covering the full API of both MPI and
 \texttt{pthread}
 with only 20 action types
 dramatically simplifies the definition of the dependence relation
and is a serious argument over working directly at the API level (like
in~\cite{OutoforderDPORGanesh}).

\section{Making Verification Efficient and Practical}
\label{sec:contrib}

This section details our two contributions, both being adaptations of ODPOR.
In the first subsection we propose an out-of-order traversal of the reduced LTS aimed at 
improving the efficiency of ODPOR. In the second subsection,
we further adapt ODPOR to better explain bugs exhibited during the traversal.

\subsection{Directed Verification}

% \color{blue}
% \begin{itemize}
% \item Parcours en profondeur explore exhaustivement les branches inutiles\\
%   Solution: exploration randomisee permettant d'eviter les vallees
% \item Pb de l'attente active: y'en a plein en MPI et ca piege un parcours en profondeur naif.\\
%  La solution: guidage en bout de branche pour defavoriser les branches de non progression
% \item The basic ODPOR is adapted to ensure that this adaptation does not hinder the soundness nor the optimality of the exploration.
%   \item on ne parle pas trop des autres algos (SDPOR etc) car on veut pas les comparer dans cet article
% \end{itemize}
% \color{black}

In the last decade, DPOR variants have been developed with the objective of either reaching optimality,
or improving the efficiency in a complete but sub-optimal exploration of the reduced LTS. 
However, for real size programs, even reduced LTS may be too large to be explored exhaustively.
The challenge that we adress is then to increase the chance to quickly find bugs when they exist,
while preserving  optimality  and  completeness in the absence of bug.

We first notice that one of the main pitfalls of DPOR techniques, which may hamper their ability  to quickly  discover bugs,  lays in their depth-first search nature.
Indeed, depending on the first explored sequence, 
there is a risk that algorithms start in a region of the state space and spend a huge
amount of time unsuccessfuly trying to find bugs there, while a
single try in a different area would perhaps immediately yield an answer. This problem is
examplified by slightly modifying the program of Fig.~\ref{lst:pattern} as follows.

\begin{example}
  Let us consider the program composed of the same $P_1$ and $P_2$, ending with the same
  actor $P_3$, but with some amount of synchronizations added in between.
  If (O)DPOR  first explores $P_1\cdot P_2$, it will then have to explore all its continuations
  before considering a (possibly faulty) run starting with $P_2\cdot P_1$.
\end{example}

We thus want to maximize the likelihood of detecting undesired behaviors in real size applications that
are too large even for reduction techniques to terminate.
The idea behind our random-first search
ODPOR (RFS ODPOR --- presented in Algo.~\ref{alg:BeFS_ODPOR}) is then to abandon the
depth-first search nature of ODPOR.
After encountering a maximal execution and computing reversible races, ODPOR  would backtrack to a state of this execution with non-empty wakeup tree,
\ie~from which some unexplored Mazurkiewicz trace remains.
Instead, in RFS ODPOR we authorize to jump to any other state with non-empty wakeup tree.
This gives much more flexibility to the algorithm in the order in which the reduced state space is built.
We could use some heuristics to select the best candidate to continue the exploration. %, thus the term ``best'' in the name of BeFS ODPOR.

As we explain now, since ODPOR is tight to the DFS traversal, the modifications are not immediate.
We detail the changes required by the algorithm while preserving soundness and optimality.

\begin{algorithm}[thb]
  \caption{RFS ODPOR($s_0$)}\label{alg:BeFS_ODPOR}
  \begin{algorithmic}[1]
    \State initialize $\expheads$ with the empty sequence $\emptyseq$;\label{lst:line:initial_expheads}
    \State choose some $t\in\enabled(s_0)$;
    \State initialize $\wut(\emptyseq)$ with $t$;
    \While{$\expheads\neq\emptyset$}
      \State choose $E\in\expheads$;
      \State choose $p$ among nodes of height one in $\wut(E)$;\label{lst:line:choose_p}
      \State $\done(E).\add(p)$;
      \State add $p$ as a child of $E$ in $\tree$;
      \State $\sleep(E\cdot p) := \{q\in\sleep(E)\cup\done(E)~|~ (p,q)\notin\mathcal{D}\}$;
      \State move the subtree of $\wut(E)$ rooted after $p$ to $\wut(E\cdot p)$;
      \If{$\wut(E)$ is empty}
        \State remove $E$ from $\expheads$;\label{lst:line:expheads_remove}
      \EndIf
      \If{$E\cdot p$ is maximal}\label{lst:line:is_enabled_empty}
        \ForAll{$i,j\in\dom(E\cdot p)$ \textbf{such that} $i\precsim_{E\cdot p}j$}
          \State $E' := \sub(E, E_{i-1})$;
          \State $v := \notdep(E_i, E).\proc(E_j)$;
          \If{$(sleep(E')\cup\done(E')_{<\proc(E_i)})\cap\weakinitial{E'}(v) = \emptyset$}\label{lst:line:intersection_wi}
            \State $E'' := \tree[E'].\mathit{insert}(v)$;\label{lst:line:fwut_reversible_race_update}
            \State $\expheads.\add(E'')$;
          \EndIf
        \EndFor
        \State $\mathit{GarbageCollect}(E\cdot p)$;
      \Else
        \If{$\wut(E\cdot p)$ is empty}
          \State choose $p'\in\enabled(\last(E\cdot p))\setminus\sleep(E\cdot p)$;\label{lst:line:p'}
          \State initialize $\wut(E\cdot p)$ with $p'$;
        \EndIf
        \State $\expheads.\add(E.p)$;\label{lst:line:add_Ep_expheads}
      \EndIf
    \EndWhile
  \end{algorithmic}
\end{algorithm}

\begin{algorithm}[thb]
  \caption{$\tree[E].\mathit{insert}(v)$}\label{alg:tree_insert}
  \begin{algorithmic}[1]
    \ForAll{$p$ \textbf{in order of} $\done(E)$}
      \If{$p \in \weakinitial{E}(v)$}
        \State \Return $\tree[E\cdot p].\mathit{insert}(v \setminus p)$;
      \EndIf
    \EndFor
    \State insert $v$ in $\wut(E)$;
    \State \Return $E$;     
  \end{algorithmic}
\end{algorithm}

\paragraph{$\wut$ and $\tree$:} as a DFS algorithm, the ODPOR algorithm only stored the stack of currently explored abstract states.
Each such state is populated with a wakeup tree, eventually grown by sequences allowing to reverse races at this state. Wakeup trees are later
explored when backtracking and then forgotten.
This raises an issue in random-first order since starting to unfold wakeup trees before having
fully populating them could lead to exploring a same trace twice, thus loosing optimality.

We store currently explored executions in a tree-like structure $\tree$,  as a  counterpart to the call stack of the ODPOR DFS.
It helps us traverse the state space at will and preserves optimality  by its combination  with wakeup trees. 
When a race is detected (line~\ref{lst:line:intersection_wi} of RFS ODPOR), a sequence $v$ is inserted at a node $E'$ in $\tree$, as described by Algorithm~\ref{alg:tree_insert}.
The insertion generalizes the insertion in the $\wut$ to the $\tree$.
It finds the leftmost (in the exploration order) and longest non-contradictory prefix in $\tree$, and continues to the current $\wut$ if necessary:
when a weak initial $p$ of $v$ is found in the current node of the tree $E$, a recursive call at node $E\cdot p$ is done with sequence $v\setminus p$, \ie~$v$ where
the first occurence of $p$ in $v$, if any, is removed;
when the algorithm fails to find such an action, it inserts $v$ in the $\wut$ of the current node $E$, and returns $E$ so that ODPOR knows where to explore later.

\paragraph{Sleep sets:} similar to ODPOR, we use sleep sets to
memorize transitions that should not be taken by future explorations, because they start sequences  equivalent to already explored ones.
In RFS ODPOR, we would need to update sleep sets as soon as some children state is explored, not only when a left subtree is fully explored like in ODPOR.
Unfortunately, directly adding those to the sleep set could result in missed executions, as illustrated by the following example.

\begin{example}
\label{ex:synchro}
Consider  an execution $E$ leading to state $s$, from which we first explore $p_1$.
Suppose that this exploration detects a race leading to insert a sequence starting with $p_2$ in $\wut(E)$.
The algorithm then decides to explore that race without completing the exploration of the
subtree $E\cdot p_1$, hence we need to add $p_1$ to the sleep set so that explorations
after $E\cdot p_2$ do not execute $p_1$ right away.
Now when we go back to $E\cdot p_1$ and discover another race after $E$ starting with $p_2$, if $p_2$ was already in the
sleep set, the race would not be inserted in the $\wut$.
\end{example}

To tackle this issue, we must add some order information to the sleep sets. 
In fact, we separate each sleep set into an inherited part, called
$\sleep(E)$ in the algorithm, and the set of processes explored from $E$ ordered according to insertion time, $\done(E)$.
When checking for a \emph{weak initial} at line~\ref{lst:line:intersection_wi}, we need to intersect with the union of $\sleep(E')$ and
those elements of $\done(E')$ inserted before the considered one for the race, denoted $\done(E')_{<\proc(E_i)}$.

\paragraph{Memory management:} 
we implement a procedure $\mathit{GarbageCollect}(E)$ which keeps in $\tree$ only nodes that could be useful in future explorations.
Indeed, due to the ordering of $\done(E)$, only explorations on the left of $E$ may impact $\wut(E)$.
Thus we need to detect when explored nodes cannot be impacted anymore and remove them. 
A leftmost leaf explored in $\tree$ can thus be removed, which triggers $\mathit{GarbageCollect(E)}$. 
Recursively,  a parent node $E'$  with no remaining  child in $\tree$ and in  $\wut(E')$ is removed.
When a parent node with children in $\tree$ is found, the procedure recursively tries to close its remaining leftmost leaf. 
The recursion stops when called from a node with non-empty $\wut$, since it will be triggered again in a future exploration of a leaf in that $\wut$. RFS ODPOR memory comsuption remains higher than its DFS counterpart as we need to
keep in memory a subtree instead of a single stack. However, thanks to the $\mathit{GarbageCollect}$ procedure, the
random choice of an element in $\expheads$ can be biased so that the total memory consumption remains under a given
threshold. In fact, if the threshold is nearly reached, the exploration can be biased to leftmost nodes in order
to release memory.

\begin{theorem}
  The RFS ODPOR algorithm is sound and optimal in the sense that its explores a unique
  execution per Mazurkiewicz's trace and explores no sleep-set-blocked execution.
\end{theorem}

\paragraph{Sketch of proof}
For the following proof we consider that $\mathit{GarbageCollect}(E)$ is disabled so that $\tree$ contains all explored sequences,
Considering $\tree$ at the end of the algorithm, proving soundness boils down to proving that it contains all Mazurkiewicz trace. 
By contradiction, let us suppose  that there exists some execution $E$ uncovered by $\tree$, \ie $\forall E'\in\tree, E\notin [E']_{\simeq}$.
Consider $\lgp_E$ the longest sequence equivalent with a prefix of $E$ which has an equivalent execution in $\tree$.
Formaly, $\lgp_E$ is the longest sequence such that there exists $v, v' \in \mathcal{T}^*$ and $E' \in\tree$ satisfying $E \simeq \lgp_E\cdot v$ and $E'\cdot v' \simeq \lgp_E\cdot v'$.
Among such sequences $E$, consider now the one with $\lgp_E$ of maximal length.
By definition of $\lgp_E$, there exists some $E'$ equivalent to $\lgp_E$ in $\tree$, such that after $E'$, RFS ODPOR explored some transitions $t_1,\dots,t_k$,
but did not explore a transition $t$ that would have led to $E$. In particular, for each $i$, $t_i \notin \weakinitial{E'}(v)$ and there is a transition
$t'_i$ in $v$ such that $D(t_i, t'_i)$,  and each transition $v_1,\dots,v_l$ in $v$ before $t'_i$ is independent with $t_i$. Hence,
due to independences, $E'\cdot v_1 \cdots v_lt_it'_i$ is a possible prefix of a trace of the programm. If the RFS ODPOR explored
that prefix at some point, the race between $t_i$ and $t'_i$ would lead to exploring their inversion. Therefore this prefix must not
have been explored. But $|\lgp_{E'\cdot v_1 \cdots v_lt_it'_i}| \geq | E' \cdot t_i| > |\lgp_E|$  contradicts the fact that $E$ is maximal for
the length of $\lgp_E$.

To prove optimality, we show that ``$\tree \cup \{\wut(E) | E\in\tree\}$ is a wakeup tree'' is an invariant for RFS ODPOR, which is similar to the proof of optimality for ODPOR.
The invariant is preserved thanks to the way insertion is done inside $\tree$ and its prolongation in the $\wut$. In
particular, respecting the order in $\done$ is crucial for verifying that if $u.p\prec u.w$ then $p\notin\weakinitial{E.u}(w)$.
Once the invariant is shown, if we suppose that two sequences in $\tree$ are equivalent, we reach a contradiction with the properties of wakeup trees.

\paragraph{Example of use: Busy-waiting.} Another interest of RFS ODPOR is its ability to
tackle programs that use busy-waiting. This classical pattern in asynchronous parallel
programming allows a process to perform computations while waiting for another process,
typically through a while loop guarded by a test on a reception or a condition variable.
In its extreme form, this pattern consists of an empty loop guarded by a test on an
asynchronous action. This is commonly used in HPC to reduce the latency by ensuring that
the communication thread remains executed on this empty loop rather than descheduled by
the operating system. That way, incoming messages get handled immediately.

Unfortunately, this construct poses a challenge to verification. 
Indeed, the application has an unbounded behavior consisting of an infinite number of tests.
However, this behaviors cannot arise in real systems, as it would imply that the process
in a busy-waiting loop executes infinitely often without the other processes to execute at
all. The natural fairness provided by real systems without failures prevents 
such behaviors. A classical DFS ODPOR could find its way out of the infinite loop by
randomly trying other processes once in a while, but it may waste time in the recursive
exploration of nearby subtrees that also have that infinite loop behavior. On the other
hand, RFS ODPOR allows the exploration to backtrack in a completely different area of the
code where the busy-waiting never occurs, hence spending more time effectively looking for
a bug.

In~\cite{Awamoche}, the authors propose a specific answer to this problem in the case of
spin loops for weak memory models. They transform busy-waiting into \texttt{assume}
statements (as explained in~\cite{DPOR_spinloop}) and perform important computations to
determine whether a given execution could be blocked before effectively visiting it. This
solution is not possible in our case since it requires access to variable values, which is
too costly in our context.

Other works propose to use techniques based on randomization to find bugs. These
techniques goes from randomizing the scheduler~\cite{Musuvathi} to injecting randomly
generated Byzantine behaviors~\cite{RandomByzantine}. Using random in those approaches
helps cover a certain part of the program state space and returns results with certain
probabilistic guarantees. Those differ from our work in the sense that randomness only
helps us to order the exploration in the hope of finding a bug faster. In particular, the
exploration remains sound and will give a certain result, not a probabilistic one.

\subsection{Counter-Example Explainability}
% \color{blue}
% \begin{itemize}
% \item   L'interet majeur du MC, c'est de donner le contre-exemple a l'utilisateur. Pb: les traces sont parfois indigestes.
% \item Root cause analysis is a large problem, and Mc~SimGrid helps identifying the critical section.
% \item Donner l'intuition de quoi on parle, et de comment la recherce se passe
% \item  Parler limites, et extensions futures ?
% \end{itemize}
% \color{black}

In this section, we focus on refinning counter-examples found by previous algorithm. 

Counter-examples are  helpful for developers to  understand the flaws and better correct them.
Surprisingly, few work deal with  explanability of counter-examples~\cite{kaleeswaran2022systematic}.
Moreover, most work related to model checking need the full state space to identify actions that cause bugs (see~\eg~\cite{BarbonLS21}). 
These techniques cannot be applied in stateless model checking, the transition system of the application is never explicitly built.
We then propose to perform some sort of root cause analysis as an extension of the ODPOR algorithm, and show how it also adapts to our RFS ODPOR.
Our notion of critical transition is close to the notion of neighbourhood of~\cite{BarbonLS21} but contrary to their solution, our algorithm
can be adaoted to ODPOR.

The principle is the following.
In a run $E$ that fails a property, we call  \newdef{critical transition} (CT) the last transitions $\gamma$ after which the bug inevitably occurs.
Formaly, let $\varphi$ be a safety property, $E$ such that $E \nvDash \varphi$. 
The \emph{critical transition of} $E$,  is the unique transition $\gamma = (s, a, s')\in E$ such that,
calling $E'$ the prefix of $E$ ending in $s$, 
all maximal continuations of $E'\cdot \gamma$ violate $\varphi$
and there exists a maximal continuation of $E'$  satisfying $\varphi$.

\begin{example}
  Let us recall that example~\ref{lst:pattern} contains a deadlock, and that a possible counterexample is
  $P_2\cdot P_1\cdot P_3\cdot P_3$. Now, if actors 1 and 2 have more code to execute, the counterexamples can
  become much longer. In fact, they can execute anything that does not interact with actor 3, and the deadlock will
  only be found when there are done. On the other hand, any execution starting with the action of actor 2 will
  lead to the deadlock. This is the critical transition to this bug, no matter what the other actions of actor 1
  and 2 are.
\end{example}

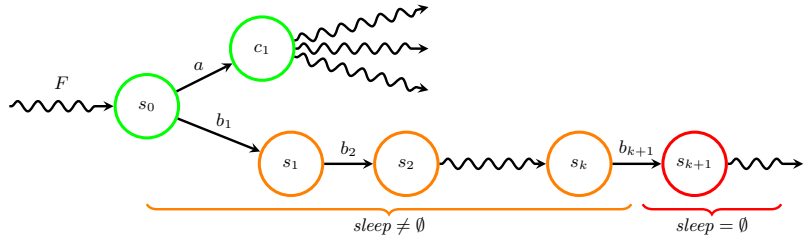
\begin{figure*}[tbh]
  \centering
    \resizebox{.9\textwidth}{!}{%
  \begin{tikzpicture}[]
    % Define the nodes
    \node[] (O) at (-2.5, -0.5) {};
    \node[shape=circle, draw=green, minimum size = 11mm, line width=1.5] (A) at (0,-0.5)   {$s_0$};
    \node[shape=circle, draw=orange, minimum size = 11mm, line width=1.5] (B0) at (2.5,-1.5) {$s_1$};
    \node[shape=circle, draw=orange, minimum size = 11mm, line width=1.5] (B1) at (4.5,-1.5) {$s_2$};
    \node[shape=circle, draw=orange, minimum size = 11mm, line width=1.5] (B2) at (7.5,-1.5) {$s_k$};
    \node[shape=circle, draw=red, minimum size = 11mm, line width=1.5] (B3) at (9.5,-1.5) {$s_{k+1}$};
    \node[] (B4) at (11.5,-1.5) {};
    
    \draw[-stealth, line width=1.2, decorate, decoration = {
      snake,  
      pre length=1pt,post length=5pt,% <-- for better looking of arrow,
    }] (O) -- node[above, yshift=0.5em]{$F$}(A);
    \draw[-stealth,line width=1.2] (A) --node[above, xshift=0.3em]{$b_1$}(B0);
    \draw[-stealth,line width=1.2] (B0) --node[above]{$b_2$}(B1);
    \draw[-stealth, line width=1.2, decorate, decoration = {
      snake,  
      pre length=1pt,post length=5pt,% <-- for better looking of arrow,
    }] (B1) -- (B2);
    \draw[-stealth,line width=1.2] (B2) --node[above]{$b_{k+1}$}(B3);
    \draw[-stealth,line width=1.2, decorate, decoration = {
      snake,  
      pre length=1pt,post length=5pt,% <-- for better looking of arrow,
    }] (B3) --(B4);

    \node[shape=circle, draw=green, minimum size = 11mm, line width=1.5] (C0) at (2,0.5) {$c_1$};
    \node[] (C1) at (5, -0.25) {};
    \node[] (C2) at (5, 0.5) {};
    \node[] (C3) at (5, 1.25) {};

    \draw[-stealth,line width=1.2] (A) --node[above, xshift=-0.3em]{$a$}(C0);
    \draw[-stealth,line width=1.2, decorate, decoration = {
      snake,  
      pre length=1pt,post length=5pt,% <-- for better looking of arrow,
    }] (C0) --(C1);
    \draw[-stealth,line width=1.2, decorate, decoration = {
      snake,  
      pre length=1pt,post length=5pt,% <-- for better looking of arrow,
    }] (C0) --(C2);
    \draw[-stealth,line width=1.2, decorate, decoration = {
      snake,  
      pre length=1pt,post length=5pt,% <-- for better looking of arrow,
    }] (C0) --(C3);

    \draw [line width=1.2, decorate, decoration={mirror,brace,amplitude=5pt}, draw=orange]
    (0,-2.2) -- node[below, yshift=-0.3em] {$\sleep \neq\emptyset$} (8.4,-2.2) ;
    \draw [ line width=1.2, decorate, decoration={mirror,brace,amplitude=5pt}, draw=red]
    (8.6,-2.2) -- node[below, yshift=-0.3em] {$\sleep = \emptyset$} (11,-2.2) ;

  \end{tikzpicture}
  }
  \caption{An exploration tree at some point during CTF+ODPOR algorithm.}
  \label{fig:CTF}
\end{figure*}

We propose an algorithm to find the CT that takes advantage of the
reduction operated by ODPOR. 
We illustrate the principles by Fig.~\ref{fig:CTF}.
After finding a first faulty execution $E$, we switch to the CT mode and keep exploring the reduced state space until finding a correct execution.
We then need to decide whether transitions in $E$, some of them with unexplored siblings, lead to correct or incorrect executions.
Let $E=F\cdot b_1 \cdots b_n$  with $F$ the prefix from which all alternative traces to $E$ (the subtree rooted in $c_1$) have been explored and are correct, by definition.
The critical transition is necessarily among $b_1,\ldots, b_n$.
The set of states $\{s_1,\ldots,s_n\}$ can always be partitionned into older ones $S_1=\{s_1, \ldots, s_k\}$ with non-empty sleep sets, and
the younger ones $S_2=\{s_{k+1}, \ldots, s_n\}$ with empty sleep sets. 
We can prove (see Appendix) that states in $S_1$ belong to correct executions, so that the critical section lead to a state in $S_2$.
Starting from $s_n$, the ODPOR algorithm then backtracks in $S_2$, using wakeup trees to explore alternatives, until finding a correct execution. 
The adaptation to RFS ODPOR is similar but requires special attention when the underlying program crashes as explained in the appendix.  

It is worth noting that the CT only identifies the last action that leads to an issue.
For instance, in the case of a four-actor deadlock, where $P_1$ and $P_2$ are blocking
each other while $P_3$ and $P_4$ are doing the same, the CT is not enough. It captures
only the last occurring pair of deadlock. Furthermore, one can look at the causal past of
the CT, \ie the set of transitions executed before the CT, causally related to it. This is
interesting since in most cases, it allows replicating the found bug with a much smaller
counterexample. But again, the same four-actor deadlock shows this is not enough, and
that, in the general case, the causal past does not fully reproduce the bug.  In the
future, we plan to refine the notion in order to tackle those cases where bugs are caused
by apparently unrelated actions. This is mandatory in order to fully take advantage of the
information conveyed by the CT.

% \todo[inline]{Voir la section related work du papier de Salaun, en particulier :
  
%   [20] G. Gosler and D. L. M{\'{e}}tayer, “A general trace-based framework of
% logical causality,” in Proc. Int. Workshop Formal Aspects Component
% Softw., 2013, pp. 157–173.

% [21] A. Beer, S. Heidinger, U. Kuhne, F. Leitner-Fischer, and S. Leue,
% “Symbolic causality checking using bounded model checking,” in
% Proc. Int. SPIN Workshop Model Checking Softw., 2015, pp. 203–221.

% [22] H. Jin, K. Ravi, and F. Somenzi, “Fate and free will in error traces,”
% in Proc. Int. Conf. Tools Algorithms Construction Anal. Syst., 2002,
% pp. 445–459.

% [23] T. Ball, M. Naik, and S. K. Rajamani, “From symptom to cause:
% Localizing errors in counterexample traces,” in Proc. 30th ACM SIG-
% PLAN-SIGACT Symp. Principles Program. Lang., 2003, pp. 97–105
% }

\section{Evaluation}\label{sec:xp}
% \color{blue}
% \begin{itemize}
% \item MBI?
% \item CorrBench pour montrer l'interet de l'evitement d'attente active et
%       celui du multifork
% \end{itemize}
% \color{black}

We implemented a prototype of our algorithm in Mc~SimGrid v3.36.1 to serve as a proof of
concept. All experiments were run on Intel Xeon E5-2630 v3 at max 2.40GHz. Programs were
compiled using \texttt{gcc}-10.2.1 with \texttt{-O3} parameter.
%The Python runners and all execution parameters are available in the artifacts.

\subsection{Evaluating RFS overhead}
\label{sec:xp_overhead}

In this first experiment, we compare the performance of our implementation of RFS with two
state-of-the-art software model checker Nidhugg~\cite{ODPOR-POPL14} and DPU~\cite{DPU}
over examples taken from the latter evaluation. Nidhugg is an implementation of
source-DPOR, a slightly different version of ODPOR that can encounter SSB executions.  It
is tuned to search pthread code for data races under weak memory models. DPU is a proof of
concept implementing unfolding DPOR, an optimal reduction algorithm based on partial event
systems. DPU can only verify mutexes operations. The benchmark against which we are
testing is composed of examples taken from the SV-COMP~\cite{SVCOMP17} repository. These
examples are bug-free, and scalable in terms of number of processes.

\begin{table}[htb]
\begin{center}
%\begin{footnotesize}
\begin{tabular}{||cc|rr|rr|rr||}
% This table has been automatically generated by runtable1.sh

% Benchmark                                                        DUP (optimal)         Nidhugg                           McSimgrid
% --------------------------------------------------------------- --------------------- --------------------------------- -------------------------
% Name                       LOC       Thrs      Confs     Events      Time        Mem        Time        Mem       SSBs        Time        States
  \hline
  \multicolumn{2}{||l|}{Benchmark} & \multicolumn{2}{l|}{DPU} & \multicolumn{2}{l|}{Nidhugg} &  \multicolumn{2}{c||}{McSimGrid RFS} \\
  Name         &   Traces &     Time   &      Mem  &   Time   &      Mem  &     Time   &      Mem \\
  \hline
  DISP(5,3)   &       1482 &      0.629 &      55M  &    6.314 &      65M &          2.665 &      50M   \\  
  DISP(5,4)   &      15282 &      6.285 &     135M  &   65.034 &      65M &         28.445 &     442M   \\      
  DISP(5,5)   &     151032 &    203.785 &     973M  &       TO &      65M &      289.361   &    4191M   \\ 
  DISP(5,6)   &            &        ERR &    1016M  &       TO &      65M &             TO &   10798M   \\ 
  MPAT(5)     &       3840 &      1.860 &      80M  &    1.203 &      64M &         7.210  &     144M   \\ 
  MPAT(6)     &      46080 &     51.283 &     420M  &   16.273 &      64M &      94.917    &    1770M   \\ 
  MPAT(7)     &     645120 &         TO &    1553M  &  255.109 &      64M &             TO &   12951M   \\ 
  MPAT(8)     &            &         TO &    1603M  &       TO &      64M &              TO &   17055M   \\ 
  MPC(3,5)    &       2958 &      0.937 &      61M  &   37.662 &      65M &         4.709 &      75M   \\ 
  MPC(4,5)    &     313683 &        ERR &      63M  &       TO &      65M &    521.814 &    6383M   \\ 
  MPC(5,5)    &            &         TO &    1344M  &       TO &      65M &            TO &   15693M   \\ 
  PI(7)       &       5040 &      1.950 &      66M  &      ERR &      66M &         6.019 &      71M   \\ 
  PI(8)       &      40320 &     28.748 &     273M  &      ERR &      66M &      50.453 &     555M   \\ 
  PI(9)       &     362880 &         TO &    1128M  &      ERR &      65M &      505.053 &    5172M   \\ 
  POKE(8)     &       3700 &      1.934 &      99M  &  146.232 &      65M &       11.427 &     186M   \\ 
  POKE(9)     &       5332 &      2.913 &     124M  &  458.337 &      65M &        18.004 &     292M   \\ 
  POKE(10)    &       7384 &      4.479 &     152M  &       TO &      64M &      26.799 &     446M   \\ 
  POKE(11)    &       9904 &      6.674 &     193M  &       TO &      65M &       38.861 &     656M   \\ 
  POKE(12)    &      12940 &      9.969 &     242M  &       TO &      65M &        53.417 &     874M   \\
  POKE(13)    &      16540 &     14.506 &     310M  &      ERR &      64M &        72.032 &    1214M   \\
  \hline
\end{tabular}

%\end{footnotesize}
\caption{Performance comparison for exhaustive exploration. TO means the execution did not finish in 10 minutes while ERR reports a
  runtime error.}\label{tab:tab1}
\end{center}
\end{table}

The results are presented in Table~\ref{tab:tab1}. Experiments have a ten minutes timeout. For each run, we report the time in seconds and
the peak of memory consumption by the applications. Nidhugg can not run on example PI as it involves C primitives it does not cover.
Overall, we observe that our RFS implementation performs in the same order of magnitude as the state-of-the-art tools regarding  time.
On the other hand, the memory consumption is higher.
This was intended as the algorithm may require to keep a tree rather than only a stack in memory.
As a conclusion, the overhead of the RFS approach is  not a bottleneck for these examples.
If memory consumption becomes an issue on bigger examples, it is still possible to tailor the order of exploration.
For example when the memory usage reaches a given threshold, the exploration could focus on leftmost states so that the $\mathit{GarbageCollect}$
procedure reclaims memory. 

\subsection{Evaluating RFS benefits for bug finding}

To evaluate the benefits of changing the order of exploration, we ran our algorithm over three bugged codes that can be scaled at will
either in the number of synchronizations or in the number of processes:
\begin{itemize}
\item \textit{MPI-any} is the MPI code of the program in Fig.~\ref{lst:pattern}
  The scaling factor is the number of \textit{rendez\-vous}
  (\texttt{MPI\_Barrier}) between the sendings and the receptions. Those synchronizations
  consist of broadcasts.
%  messages sent from every process to every other process.
  Hence, adding one
  \textit{rendez-vous} effectively doubles the number of interleaving to consider.
\item \textit{philosophers-mutex-deadlock} is an implementation of the dining
  philosophers using one mutex per fork, scaling with the number of
  philosophers. A deadlock occurs if all philosopher pick their 
  first fork before a philosopher considers its second fork.
\item \textit{philosophers-semaphore-deadlock} is a variant of the same program,
%  another implementation of the dining   philosophers,
  with an additional semaphore restricting the number of philosophers allowed
  to pick forks
  %  , that are still protected by mutexes.
  A deadlock occurs since the initial number of
  tokens in the semaphore equals the number of philosophers.
\end{itemize}

\begin{figure*}[hp]
  \centering
  \begin{subfigure}[]{\textwidth}
    \centering
    \includegraphics[width=\textwidth]{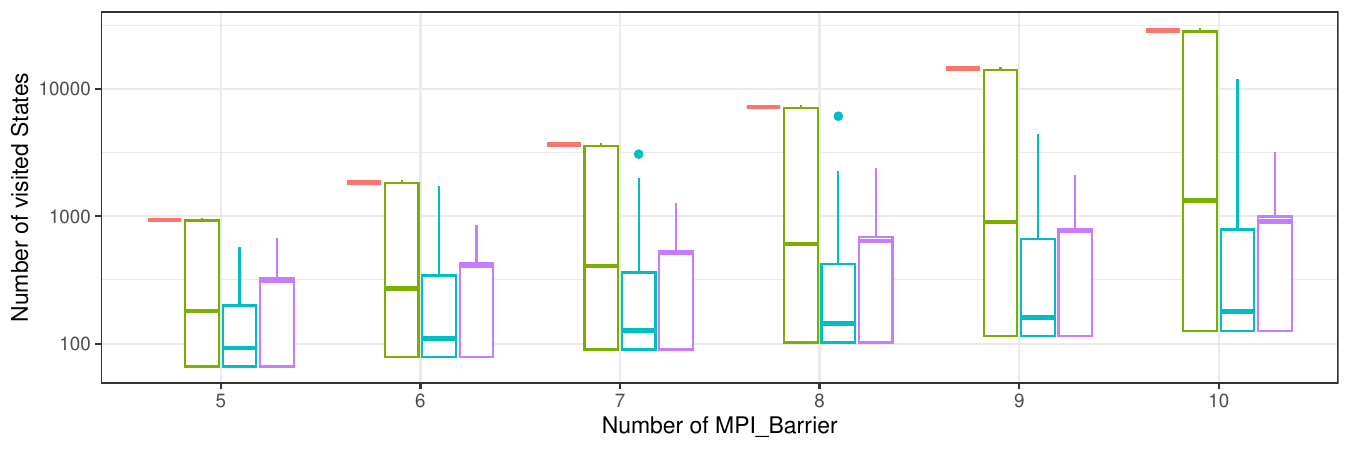}
    \caption{\textit{MPI-any}}
  \end{subfigure}
  \begin{subfigure}[]{\textwidth}
    \includegraphics[width=\textwidth]{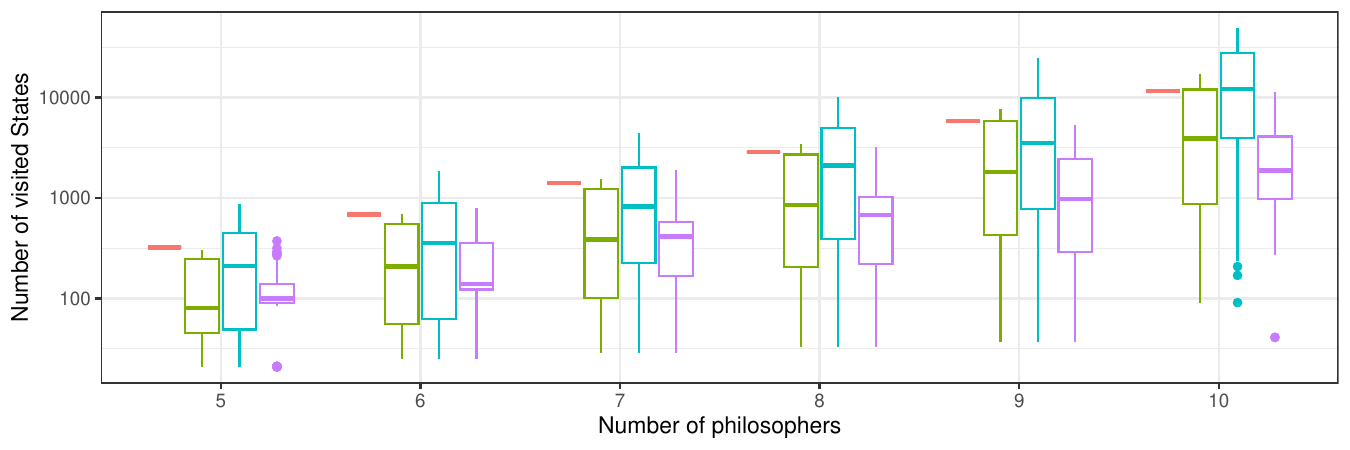}
    \caption{\textit{philosophers-mutex-deadlock}}
  \end{subfigure}
  \begin{subfigure}[]{\textwidth}
    \includegraphics[width=\textwidth]{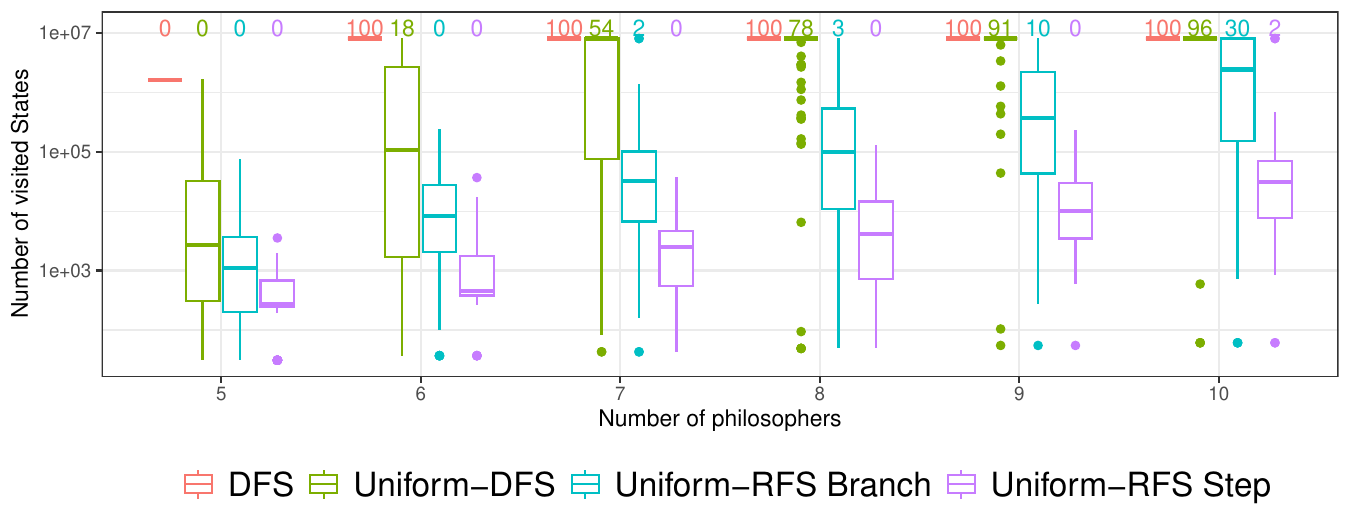}
    \caption{\textit{philosophers-semaphore-deadlock}}
  \end{subfigure}
  \caption{Number of states explored before finding the deadlock at various scales.}
  \label{fig:results}
\end{figure*}

We ran four variants of ODPOR over these examples: DFS, Uniform-DFS, Uniform-RFS
Branch and Uniform-RFS Step. DFS is the classical ODPOR algorithm
from~\cite{ODPOR-POPL14} running in an arbitrary depth-first search manner. When
confronted to a choice, it picks the process with the smaller identifier.
The Uniform-DFS variant resolves those choices with a uniform random pick.
The last two variants both resolve choices with a uniform pick:
the ``Step'' variant picks any element in $\expheads$ at each step,
while the ``Branch'' variant completes a branch before choosing any element in $\expheads$.
%In future work, heuristics could be proposed to select the best candidate continuation point. 
 %We ran each benchmark with an increasing scaling factor.
 %(barrier count in \textit{MPI-any}, and philosopher count for the other benchmarks).
Randomized explorations are run 100 times each, with 100 different seeds.
The results depicted in Fig.~\ref{fig:results} show the number of states explored before finding a deadlock
when the scaling factor varies.

All runs of \textit{MPI-any} (1st table) and  \textit{philosophers-mutex-deadlock} (2nd table) terminate in less than 5mn, while
for \textit{philosophers-semaphore-deadlock} (3rd table) some runs timeout after 10mn without finding a bug.
For these runs, the number of states explored before timeout is used, allowing
the box plots to depict the success rate of a given strategy for large applications.
The number of timeouted runs (over 100) is written above each box plot.
% \todo{TJ: il faut rajouter une phrase d'analyse des resultats} 

The first experiment with \textit{MPI-any} illustrates an example where the scaling is in
number of operations per actor. In this case, randomizing the original DFS algorithm with
Uniform-DFS helps keep the average time on a linear scale, but at least 25\% of the
executions still behave as badly as the worst-case scenario. On the opposite, both Step-RFS
and Branch-RFS runs 75\% of the time faster than the average for Uniform-DFS, with
Branch-RFS behaving a bit better in that example. This is because Branch-DFS has the
opportunity to explore a faulty execution each time it backtracks.

The other experiments are examples of scaling in the number of actors, with the semaphore
variant having many more interleaving to consider. On the smaller example of
\textit{philosophers-mutex-deadlock}, it is hard to conclude to any trend. By trying whole
sequence and never a few steps, Branch-RFS is in average the slowest version, while
Step-RFS tends to explore in parallel different sequences step-by-step, and faster reaches
deadlocks that are located somewhere in the middle of the state space.  The trend is more
visible on the biggest example \textit{philosophers-semaphore-deadlock}.  In that case,
for ten philosophers, only 4 executions out of 100 finished in under 10 minutes with
Unfirom-DFS while other executions visit more than 10 millions nodes without finishing.
Our solution performs better with Step-RFS finishing 98\% of the times, visiting an
average of only 12,000 nodes before finding the bug. Step-RFS does not behave like a BFS
exploration as one could expect. This is due to the high restriction the reduction has on
the state space. By extensively limiting the number of parallel branches opened at the
same time, the reduction turns a simple random step variation into a working trade-off
over the various situations. These preliminary experimental results are promising for the
efficiency of our techniques to detect bugs in real-size applications.

In all these experiments, we also evaluated the CT overhead. The worst case is when
the first explored execution leads to a fault, in which case CT must explore entirely the
sub-tree around that run while it can reuse previously explored traces otherwise.
That exploration is very fast in the two \emph{philosophers} benchmarks as the critical
transition is one of the last trace's actions, but takes more time for the first
benchmark where the critical transition is the very first action in the trace.
This is indeed the worst case for the CT algorithm, as it needs to exhaustively explore
the reduced LTS searching for a correct execution which does not exist.

\section{Conclusion and Future Work}\label{sec:cc}
% \color{blue}
% \begin{itemize}
% \item OpenMP
% \item Explicabilite, c'est tout un monde
% \end{itemize}
% \color{black}

In this paper, we proposed two adaptations of ODPOR, a state-of-the-art Dynamical Partial
Order Reduction technique, focused on bug finding and explanation, implemented in
Mc~SimGrid.  First, the RFS ODPOR variant allows arbitrary orderings of the state space
exploration. This degree of freedom shows promising results in discovering faulty
behaviors early. Second, we adapted further the ODPOR technique with a novel algorithm for
counter-example explainability, based on the notion of critical transition, which requires
minimal computation overhead. In the future, we plan to search for model specific
heuristics in order to further accelerate RFS ODPOR in the presence of faults in the
program. Thanks to its design, RFS ODPOR also paves the way for intensive parallelization
through producer-consumer patterns. Finally, we aim at refining our notion of critical
transition in order to explain counter-examples a step further.

% \section*{Acknowledgment}

% Experiments presented in this paper were carried out using the Grid'5000 testbed,
% supported by a scientific interest group hosted by Inria and including CNRS, RENATER and
% several Universities as well as other organizations (see https://www.grid5000.fr).

\bibliography{references}
\bibliographystyle{acm}

\newpage 

\appendix

\section{Critical Transition Search: more details}

We first recall what a critical transition is, before giving some details of the CT-search algorithm using ODPOR, and a necessary  adaptation
required for the RFS ODPOR algorithm.

Let $\varphi$ be a safety property, $E$ a \newdef{faulty} execution, \ie, $E \nvDash \varphi$. 
The \emph{critical transition of} $E$, %if it exists,
is the unique transition $\gamma = (s, a, s')\in E$ such that, calling $E'$ the
prefix of $E$ ending in $s$: 
\begin{itemize}
\item there exists a maximal continuation of $E'$  satisfying $\varphi$,
\item all maximal continuation of $E'\cdot \gamma$ violate $\varphi$.
\end{itemize}

Since a critical transition characterizes the last step necessary to produce a given
bug, it requires the existence of a faulty execution.
On the opposite side, in the limit case when there is no correct execution, the critical transition is the start transition.

We restart to explain the principle of the CT-search algorithm on a schematic representation given by Fig.~\ref{fig:CTF} of
the general situation encountered while running the CT algorithm.
Let $E=F\cdot b_1 \cdots b_n$ be the first incorrect execution reached during the exploration
phase, with $F$ a prefix from which all alternative traces to $E$ have been explored and
are thus correct (as well as all their equivalent executions).
In the figure, the execution in the subtree rooted in $c_1$ examplifies this: they have been explored previously and are all correct. Hence, the critical
transition is among $b_1,\ldots, b_n$.
Let the set of states $\{s_1,\ldots,s_n\}$ be
partitioned into those with a non-empty sleep set, $S_1=\{s_1, \ldots, s_k\}$, and
those with empty sleep sets $S_2=\{s_{k+1}, \ldots, s_n\}$.
This partition always has this form where both parts are contiguous: indeed $s_1,\ldots,s_n$ have no
siblings yet, hence $\sleep(s_n)\subseteq\sleep(s_{n-1})\subseteq\cdots\subseteq\sleep(s_1)$.
We perform a case analysis to prove  that states in $S_1$  are in a correct execution,
and then show how the critical section can be found among states in $S_2$: 
\begin{description}
  \item[Non-empty sleep set:] let $s_i, 1\leq i\leq k$ be a state in $S_1$ thus with a
non-empty sleep set, and let $p\in\sleep(s_i)$. By definition of sleep-sets, there
exists $e'\in\dom(F)$ \st, writing $F'= \pre(F,e')$, $p\in\enabled(F')$, $p$ has been
explored after $F'$, and $p$ commutes with each action in
$\proc(e')\cdots b_1\cdots b_{i-1}$. Intuitively, for $p$ to be in the sleep-set of
$s_i$, a previous exploration diverging from $E$ must have been visited. That
diverging exploration is of the form $F'\cdot p$, and $p$ must be independent with every
action taken since that moment. Since the exploration backtracked from $F'\cdot p$, by the
soundness of the ODPOR algorithm we can say that every trace starting by $F'\cdot p$ has
been explored. In particular, every trace starting by
$F'\cdot p \cdot \proc(e')\cdots b_1\cdots b_{i-1} \simeq F' \cdot \proc(e')\cdots
b_1\cdots b_{i-1}\cdot p$ has been explored. Since $F\cdot b_1 \cdots b_n$ is the first
incorrect execution encountered, $F' \cdot \proc(e')\cdots b_1\cdots b_{i-1}\cdot p$ must
lead to correct executions only.
Consequently, we can conclude that any state in $S_1$
belongs to a correct execution.
\item[Empty sleep set:] we now show how to search the CT in $S_2$.
When reached during backtracking, a state $s_i \in S_2$ has an empty sleep set,
thus there was no other explored execution equivalent to one continuing after $s_i$.
%every execution prefixed from  $s_i$ is non equivalent to any execution explored from another reached state. 
Therefore, when backtracking at $s_i$, the executions starting by transitions not in
the wakeup tree of $s_i$ are equivalent to some already explored ones (by soundness
of the ODPOR algorithm). Hence, if we did not find a correct execution during the CT up
to the state $s_i$, and we fully explored its wakeup tree, then there is no correct
execution from state $s_i$. The CT algorithm can backtrack one step further.
\end{description}

\paragraph{Preserving persistence after a crash}
One remaining difficulty to solve is linked to the fact that programs are really executed in Mc~SimGrid,
and as ODPOR algorithms are stateless, we never dispose of the complete transition system. Consider a state $s$ in which
$\enabled(s) = \{p_1,\ldots, p_n\}$, and such that executing $p_1$ leads to a crash of the
program. The issue is that, of course, no other $p_i$ can be executed, since after
executing $p_1$ the program execution does not exist anymore. Nevertheless, we need to
preserve the soundness of the reduction and ensure that we give the opportunity to the
algorithm to execute runs in other orders during CT-serach. In other words, when encountering a
bug, the persistence property of the programming model may be lost. To overcome this difficulty, we
need to adapt RFS ODPOR so that it keeps exploring other traces when reaching such
crashes. This adaption is close to what is proposed by the author of
ODPOR~\cite{ODPOR-POPL14} in the case of non-persistent models. The difference is that we
only adapt it for the very specific case of bugs encountered during CT, while covering
the general case requires much more adaptation.

%(see also~\cite{Faqrizal-Salaun-Formalise2022})

%\section{Implementation details}

\end{document}